\documentclass[onecollarge,final,sort&compress]{svjour2}
\usepackage[numbers]{natbib}
\journalname{Few Body Systems}
\usepackage{graphicx}
\usepackage{slashed}
\usepackage{amstext}

\begin{document}

\title{
On a new approach to meson phenomenology with the Bethe-Salpeter equation
\thanks{Presented by C.\,Popovici at Light-Cone 2014, NC State
University, Raleigh, USA, 26 - 30 May, 2014}
}

\author{Carina Popovici	\and
		Thomas Hilger	\and
		Mar\'ia G\'omez-Rocha \and
		Andreas Krassnigg 
}

\institute{C. Popovici et al. \at
           University of Graz, Institute of Physics, NAWI Graz, A-8010 Graz, Austria\\
           \email{carina.popovici@uni-graz.at}  
}

\date{Received: \today}

\maketitle

\begin{abstract}
We investigate capabilities of the effective interaction in a
rainbow-ladder truncated meson model of QCD within a covariant
Landau-gauge Bethe-Salpeter-equation approach.  Based upon past
success for the light- as well as heavy-quark domains, we discuss the
range of applicability and features of an effort with
comprehensive phenomenological claim and goals.

\end{abstract}

\section{Introduction}
\label{sec:intro}

The calculation of meson properties in the Dyson-Schwinger-Bethe-Salpeter-equation (DSBSE) approach
has enriched the theoretical hadron-physics landscape for many years. In fact, it was realized
soon after the conception of the quark picture of hadrons that a relativistic dynamical setup was
needed for a more in-depth description of the ever-growing sample of hadron-physics data. Moreover,
the phenomenological success of the quark-model hypothesis clearly indicated the convincing potential
of a covariant description of hadrons rooted in quantum chromodynamics (QCD), which is now widely
accepted to be the theory describing the strong interaction. The modern tools that displayed the
capability to achieve this goal are lattice-regularized QCD on one hand and continuum quantum field
theoretical methods on the other hand, one of which is the DSBSE approach employed here.

In a phenomenological DSBSE setup one is immediately confronted with the complexity of the infinite,
coupled system of QCD's Dyson-Schwinger equations (DSEs) \cite{Fischer:2006ub}. Thus, the straight-forward 
rainbow-ladder (RL) truncation of the coupled quark DSE and meson Bethe-Salpeter-equation (BSE) 
system rose to great popularity quickly 
\cite{Maris:2003vk} after its helpful and QCD-authentic features had been demonstrated. In particular, 
attention was drawn to relevant Ward-Takahashi identities (WTIs) such as the axial-vector WTI (AVWTI), 
see, e.\,g., \cite{Munczek:1994zz}, and its satisfaction in RL truncation
(together with the vector WTI \cite{Maskawa:1974vs,Aoki:1990eq,Kugo:1992pr,Bando:1993qy,
Munczek:1994zz,Maris:1997hd,Maris:1999bh,Maris:2000sk}),
which leads to a comprehensively veracious description of the pion and its properties \cite{Maris:1997tm}.
More precisely, a pion computed from an RL-truncated DSBSE model calculation follows the pattern
required by the Goldstone theorem in that it is massless in the chiral limit. For small finite current-quark
masses, it follows the well-known Gell-Mann--Oakes--Renner relation; in fact, 
it was even shown that a generalized version of this relation exists that is valid for all pseudoscalar mesons
regardless of their mass and level of excitation \cite{Maris:1997hd,Holl:2004fr}. 

On top of the archetypical treatment of the pion in this approach, it is not surprising that
most of the phenomenological studies that followed and which used a sophisticated model interaction
in RL truncation focused on the light-quark sector. In addition, reaching larger current quark masses
and numerically computing bound states of such quarks in a Landau-gauge calculation (for insight in
the situation in Coulomb gauge, see, e.\,g., 
\cite{Alkofer:2005ug,Popovici:2010mb,Popovici:2011yz,Popovici:2011wx,LlanesEstrada:2010bs,Cotanch:2010bq,Rocha:2009xq}) 
in Euclidean space like it is used in this approach poses numerical challenges. When these were finally 
being overcome recently, see, e.\,g., \cite{Krassnigg:2008gd,Blank:2010bp}, the model assumptions remained 
anchored  to the light-quark domain nonetheless. While this still gave reasonable results for pseudoscalar 
and vector mesons, states identified with either radial or orbital angular momentum excitations were not well described 
\cite{Bhagwat:2006py,Bhagwat:2007rj,Fischer:2009jm,Krassnigg:2009zh,Krassnigg:2010mh,Blank:2010sn,Qin:2011xq,Rojas:2014aka,Fischer:2014xha}. 
While this could be interpreted on a general footing and the conclusion could be drawn that the RL truncation
is not sufficient to provide a generally satisfying meson phenomenology and that for such satisfaction
to be achieved one needs to include corrections to this truncation or, simply speaking, a quark-gluon
vertex more complicated than the bare one, we challenge this line of thinking and attempt
a counterexample. 

More precisely, we start our version of a phenomenological QCD-model approach via the DSBSE method
in the heavy-quark domain. Generalizing on previous accomplishments \cite{Blank:2011ha}, we allow for more 
freedom in the effective interaction and test our assumption by comparing our results to the available meson
data in the bottomonium system. Herein we present a first look at the possibilities and limitations
of the present setup, as well as steps to be taken next to complete this study.

\section{Bottomonium in the DSBSE approach}
\label{sec:bottomonium}

The study of the bottomonium system in the DSBSE approach has been a part of several investigations
of meson properties. This section means to put them in perspective with respect to each other
and to the comprehensive study to follow up on the present excerpt \cite{Popovici:2014mt}. 
In the context of the present setup it is always instructive
to note that first simplified attempts at meson spectroscopy including bottomonium were already undertaken
several decades ago \cite{Munczek:1983dx} and later, under certain approximations to the quark propagators
that violated the AVWTI, in \cite{Munczek:1991jb,Jain:1993qh}, where also radial excitations were studied.
This line of work was continued by investigating corrections beyond RL truncation in a systematic
truncation scheme using a simplified model interaction \cite{Bhagwat:2004hn,Gomez-Rocha:2014vsa}.

Separable forms of the BSE kernel were employed mainly to make use of concepts along the lines of heavy-quark
effective theory and study heavy-light mesons \cite{Ivanov:1998ms}.

In later studies with a full numerical account of the quark propagators and thus an also numerical
satisfaction of the AVWTI, heavy quarks were difficult to treat with methods available at the time and so
at first efforts focussed on systems involving only light or at most charmed quarks \cite{Krassnigg:2004if}. 
Bottomonium in this context first appeared
only a couple of years ago \cite{Maris:2006ea} and soon thereafter several investigations involved bottomonium as
an important part for the study of, e.\,g., effects of the dressing of heavy quarks or various parts
of the effective interaction
\cite{Krassnigg:2009zh,Souchlas:2010st,Souchlas:2010zz,Nguyen:2010yh,Nguyen:2010yj,Nguyen:2009if}.

The most recent development regarding bottomonium in this context is given in \cite{Blank:2011ha} where
bottomonium ground-state masses and decay constants were studied to test the straight-forward 
applicability of a standard effective interaction to this system simply by adjusting one free model 
parameter and without subsequent fine-tuning of any of the model parameters, which proved to be 
successful for all ground states known experimentally.

\section{Interaction model and phenomenological setup}
\label{sec:model}

In ladder truncation the homogeneous BSE for
quark-antiquark bound states reads:
\begin{eqnarray}
\Gamma(p;P)&=&-C_F\int^\Lambda_q\!\!\!\!\mathcal{G}((p-q)^2)\; D_{\mu\nu}^f(p-q) \;
\gamma_\mu \; S(q_+) \Gamma(q;P) S(q_-)\;\gamma_\nu \;,\label{eq:bse}
\end{eqnarray}
where $\Gamma$ is the Bethe-Salpeter amplitude (BSA), $C_F=4/3$ the Casimir color factor,
$D_{\mu\nu}^f$ is the free gluon propagator, $\gamma$ is
the Dirac part of the bare quark-gluon vertex, and
$\int^\Lambda_q:=\int^\Lambda d^4q/(2\pi)^4$ represents a
translationally invariant regularization of the integral, with the
regularization scale $\Lambda$ \cite{Maris:1997tm}.
$q$ and $P$ are the
relative and total momenta of the $q\bar{q}$ state, respectively, and
the semicolon separates them as four-vector arguments of the BSA.  The (anti)quark
momenta are $q_+ = q+\eta P$ and $q_- = q- (1-\eta) P$, where $\eta
\in [0,1]$ is referred to as the momentum partitioning parameter.
We use the arbitrariness of the value of $\eta$ in our covariant
framework to set $\eta=1/2$.

The renormalized dressed quark propagator $S(p)$ is obtained from
the corresponding rainbow-truncated quark DSE
\begin{eqnarray}\label{eq:dse}
S(p)^{-1}  &=&  (i\gamma\cdot p + m_q)+  \Sigma(p)\,,\\\label{eq:selfenergy}
\Sigma(p)&=& C_F\int^\Lambda_q\!\!\!\! \mathcal{G}((p-q)^2) \; D_{\mu\nu}^f(p-q)
\;\gamma_\mu \;S(q)\; \gamma_\nu \,.
\end{eqnarray}
$\Sigma(p)$ denotes the quark self-energy, and $m_q$ is the current-quark mass; details
of the renormalization of the quark propagator can be found in
\cite{Maris:1997tm,Maris:1999nt}.

The function $\mathcal{G}$ apparent in both Eqs.~(\ref{eq:bse}) and (\ref{eq:selfenergy}) 
is the effective form of the quark-qluon interaction to go with the RL truncated model setup.
With $s:=(p-q)^2$ we employ the well-established parameterization \cite{Maris:1999nt}
\begin{equation}
\label{eq:interaction} 
\frac{{\cal G}(s)}{s} =
\frac{4\pi^2 D}{\omega^6} s\;\mathrm{e}^{-s/\omega^2}
+\frac{4\pi\;\gamma_m \pi\;\mathcal{F}(s) }{1/2 \ln
[\tau\!+\!(1\!+\!s/\Lambda_\mathrm{QCD}^2)^2]}.
\end{equation} 
This form has a perturbative limit consistent with the one-loop
renormalization group behavior of QCD. While the far infrared is not expected to
have a significant impact for our purposes \cite{Blank:2010pa}, its low and intermediate momentum ranges
include some model enhancement to provide the flexibility needed in a phenomenological
approach, e.\,g., to accommodate the correct amount of dynamical chiral symmetry breaking.
Furthermore, ${\cal F}(s)= [1 - \exp(-s/[4
m_t^2])]/s$, $m_t=0.5$~GeV, $\tau={\rm e}^2-1$, $N_f=4$,
$\Lambda_\mathrm{QCD}^{N_f=4}= 0.234\,{\rm GeV}$, and
$\gamma_m=12/(33-2N_f)$ \cite{Maris:1999nt}.

This model interaction has been used over the past years to successfully describe hadron properties, 
most prominently but not limited to the ones of pseudoscalar and vector mesons, such as electromagnetic 
properties \cite{Maris:1999bh,Maris:2005tt,Holl:2005vu,Bhagwat:2006pu,Eichmann:2007nn,Eichmann:2011vu},
strong decay widths \cite{Jarecke:2002xd,Mader:2011zf}, valence-quark distributions 
\cite{Nguyen:2011jy,Cloet:2013jya}, as well as properties at finite
temperature \cite{Maris:2000ig,Blank:2010bz}.

\section{Approach}
\label{sec:approach}

In this section we outline our strategy for obtaining a DSBSE result for 
the bottomonium spectrum that is most satisfactory in the current setup.
Notably, two differences compared to the previous study in \cite{Blank:2011ha}
appear: First, we attempt to describe the spectrum of not only ground
but also radially excited states. Second, we allow additional variation
of the model parameters in Eq.~(\ref{eq:interaction}). Keeping this in mind, 
we thus test our model effective interaction within the range specified
below with regard to the following challenges:
\begin{itemize}
\item reproduce the splittings of bottomonium ground-state masses for
the states available experimentally for $J=0,1,2$ with the same quality as already achieved
in \cite{Blank:2011ha}
\item in addition, reproduce the splitting of the ground vs.~first radially excited state in 
each channel experimentally available
\item alternatively, reproduce the splittings of all first radially excited states with
respect to each other, where experimentally available
\end{itemize}

It is important to note at this point that this is the first study with the declared goal
to successfully describe both ground and radially excited meson states in an RL-truncated 
DSBSE approach. Since it is not clear a priori that such an endeavor can be successful
even for the promising realm of heavy-quark bound states, several steps are needed to test
model assumptions and restrictions without losing track of where certain changes come from.

The original setup of Maris and Tandy \cite{Maris:1999nt} for their interaction was 
anchored in the light-quark domain and model parameters were adjusted to relevant quantities,
namely the pion mass and decay constant as well as the chiral condensate. The relevant
term in the effective interaction Eq.~(\ref{eq:interaction}) is the first one, while the
second determines the behavior of calculated results in or towards the perturbative domain. 
More precisely, the current-quark mass $m_q$ as well as the parameters $\omega$ and $D$ 
were adjusted such that light pseudoscalar and vector meson masses and decay constants were well 
described by, as it turned out, fixing the product $D\times\omega$ to $0.372$ GeV${}^3$ and varying
$\omega$ in the range $[0.3,0.5]$ GeV. In this way, the choice of $D\times\omega$ and $m_q$ effectively
defined a one-parameter model. While the calculated pseudoscalar and vector ground-state observables were 
independent of $\omega$, it was shown later that radial- and orbital-excitation properties strongly 
depend on $\omega$, even with a fixed value for $D\times\omega$, see \cite{Krassnigg:2009zh} and 
references therein. This is not surprising, since $\omega$ corresponds to an inverse range of
the intermediate-momentum (i.\,e., the long-range) part of the effective interaction and one 
would expect such a parameter to have a noticeable effect on excited but not ground states \cite{Holl:2004un}.

In \cite{Blank:2011ha} the original value for the product $D\times\omega=0.372$ was kept and $\omega$ fitted
to $\omega=0.61$ GeV to achieve excellent agreement with the experimentally known bottomonium ground states.
An equally successful description of radial excitations in addition to the ground states is not possible
without allowing both $\omega$ \emph{and} $D$ to vary \emph{independently}, which is what we have done to
arrive at the results presented here. 

\begin{figure*}
\centering
 \includegraphics[width=1.0\textwidth]{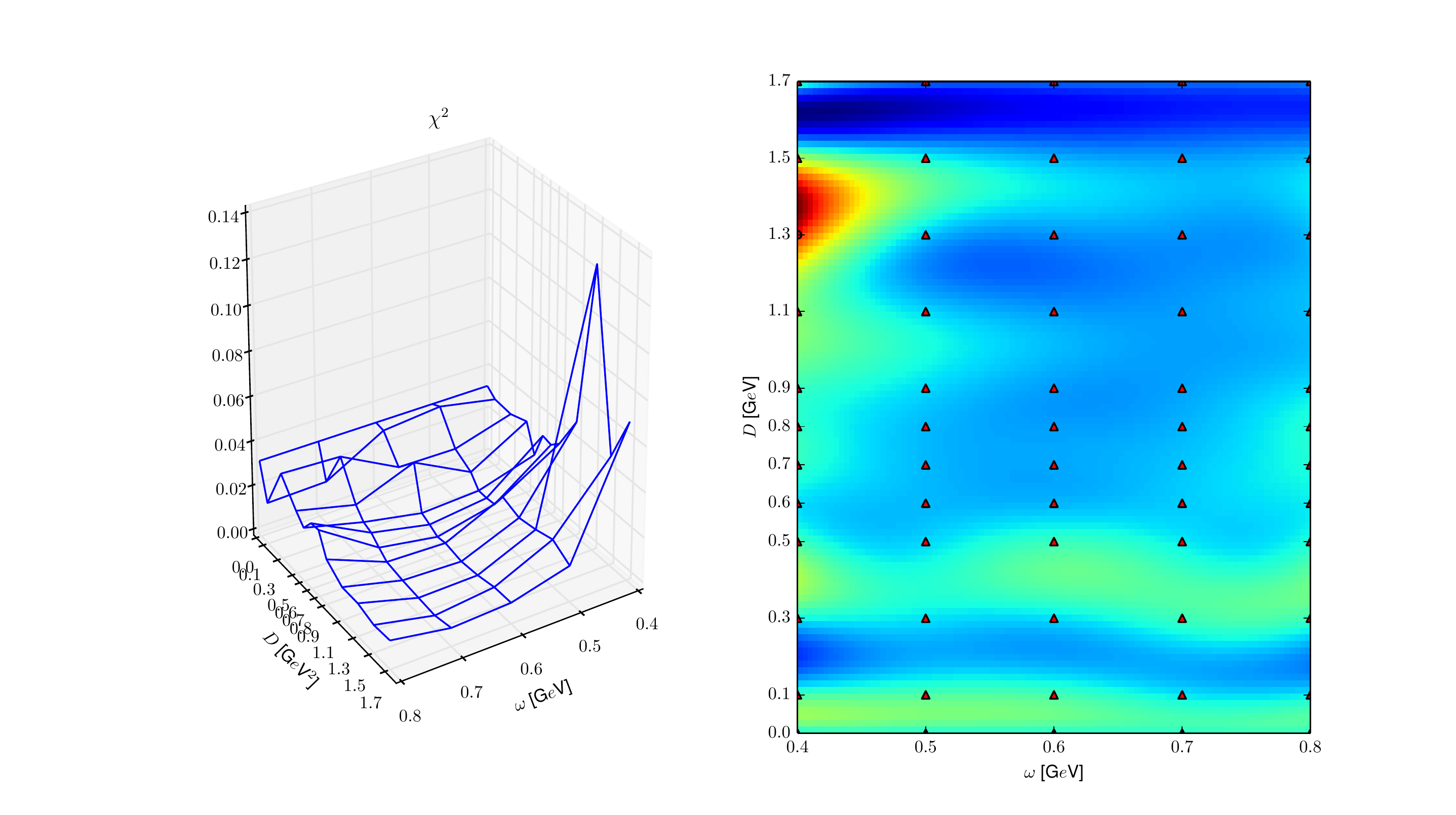}
\caption{\emph{Left panel}: $\chi^2$ for a combination of splittings calculated in the
  $[\omega, D]$ plane for the bottomonium system and compared to experimental data \cite{Beringer:1900zz}.
  \emph{Right panel}: contour plot for the same data with interpolated values in between our
  grid points (red triangles).}
\label{fig:splittings}      
\end{figure*}
More precisely, as a first step we calculate the mass-splittings among ground and excited states in the bottomonium
system for a number of values on an $\omega$-$D$ grid for a fixed value of the bottom current-quark mass and 
plotted the corresponding $\chi^2$ resulting
from our comparison with the available experimental numbers for those splittings, as shown in the left
panel of Fig.~\ref{fig:splittings}. The right panel of this figure illustrates the behavior of a spline of our
grid data.

The second step is then to use the optimal value combination of $\omega$, $D$ on our grid and adjust the 
bottom current-quark mass $m_b$ such that the experimentally known ground-state masses in the bottomonium
system are best reproduced in a least-squares fitting procedure. More concretely, the masses used for this fit
have the quantum numbers $J^{PC}=0^{-+}$, $0^{++}$, $1^{--}$, $1^{++}$, $1^{+-}$, and
$2^{++}$. The masses of the remaining states are thus predictions of the model.

\section{Results and Discussion}
\label{sec:results}

\begin{figure*}
\centering
 \includegraphics[width=0.7\textwidth,clip=true]{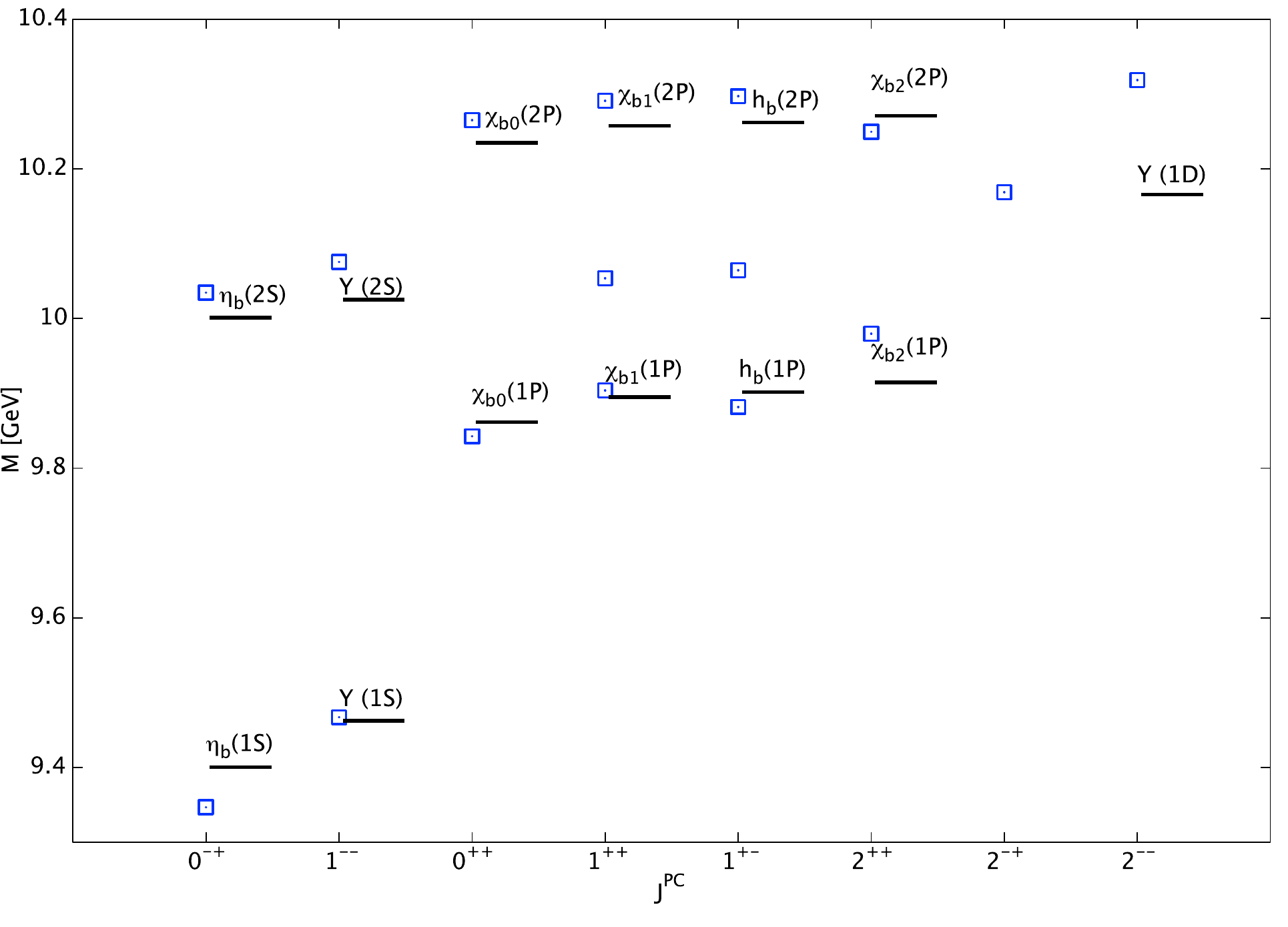}
\caption{Bottomonium spectrum: calculated (symbols) versus
  experimental (lines) data. Error bars are contained inside the symbols for each set of data.}
\label{fig:bottomonium}      
\end{figure*}

The set of various splittings among bottomonium ground and excited states
was computed for a bottom-quark mass of $m_b=3.71$ GeV (given at a renormalization point $\mu=19$ GeV) 
and is best reproduced on our grid by the combination $\omega=0.7$ GeV and $D=1.3$ GeV${}^2$.
The subsequent least-squares fit of the ground-state masses as described above yields
$m_b=3.635$ GeV; our corresponding results are depicted in Fig.~\ref{fig:bottomonium}, where we also
provide the experimental data. The agreement is surprisingly good with the exception of two ``extra'' states
that appear as calculated excitations in the $J^{PC}=1^{++}$ and $J^{PC}=1^{+-}$ channels.
Clearly, the nature of these states needs further investigation, which is carried out at the moment.
Attempts to include these states as the first radial excitations in their respective channels were unsuccessful,
which may also hint at the fact that further degrees of freedom are needed in the effective interaction
to provide an overall satisfactory description of the bottomonium system with all its excitations.
We note that, since both our numerical as well as experimental uncertainties are smaller than the respective
symbol sizes, we have not plotted error bars in Fig.~\ref{fig:bottomonium}.

\section{Conclusions}

Building on the success of a previous study of the bottomonium ground sates in an RL truncated DSBSE approach, we have provided 
the first successful combined description of ground and radially excited states for the bottomonium system by 
allowing a wider and more independent variation of the parameters in the effective model interaction.
This is the immediate consequence of the idea to anchor the effective quark-gluon interaction in the heavy-quark
domain; in addition, our ultimate goal is to provide a comprehensive description of meson spectra along the whole range 
of quark masses from bottomonium down to the chiral limit, possibly allowing the effective interaction to depend on the current
quark mass (see, e.\,g., \cite{Williams:2014iea} for recent insight regarding this topic). This might mimic effects beyond RL truncation such that a successful description of both ground-
and excited-state meson properties can be maintained also in the charmonium system and, ultimately, the light-quark sector.

\begin{acknowledgements}
We acknowledge helpful conversations with M.\,Blank.  This work was
supported by the Austrian Science Fund (FWF) under project no.\
P25121-N27.
\end{acknowledgements}

\end{document}